\documentclass{article}

\begin{document}

\title{The art of science: interview with Professor John~Archibald~Wheeler}
\author{Ji\v{r}\'{i} Bi\v{c}\'{a}k\footnote{Jiri.Bicak@.mff.cuni.cz}\vspace{1em}\\
Institute of Theoretical Physics,\\
Faculty of Mathematics and Physics, Charles University,\\
 V Hole\v{s}ovi\v{c}k\'{a}ch 2, 180\,00 Prague 8, Czech Republic
}
\maketitle

\begin{center}
\parbox{.9\textwidth}{
\emph{Published in General Relativity and Gravitation \textbf{41} (2009), 679-689, the special issue on quantum gravity, dedicated to the memory of John Archibald Wheeler.}}
\end{center}

\vspace{1em}

John Archibald Wheeler, born on July 9, 1911 in Jacksonville, Florida, died on April 13, 2008. He must be known to all readers of \emph{General Relativity and Gravitation}, if not from anything else, then at least as ``W'' in the relativists' ``biblical text'' MTW and as the physicist who coined the term ``black hole''.

By fundamental work and by many new ideas Professor Wheeler enriched different fields of modern physics, including nuclear physics, elementary particle physics, general relativity, astrophysics and cosmology. He was the teacher of such physicists as Richard Feynman, John Klauder, Charles Misner, Kip Thorne and many others. His life was deeply influenced by a close relationship with Niels Bohr before and during World War II. After the death of Albert Einstein in Princeton in 1955, Professor Wheeler became a leading personality in the general theory of relativity in Princeton---and in the whole world. He was a member of many scientific societies (the president of the American Physical Society in 1966) and the recipient of many medals and awards.

During the conference on the methods of differential geometry in physics in Warsaw in June 1976, Professor Wheeler gave an interview for the Czechoslovak Journal of Physics A (published in Czech, in contrast to the series B). After 1968, an unhappy year for Czechoslovakia, I became a member of the editorial board of this Journal (replacing Karel Kucha\v{r} who moved to Princeton following the invitation from Prof. Wheeler). The journal was quite independent of then common censorship; by means of interviews with interesting personalities we tried to keep some ties with the ``external world''. The following interview was quite enthusiastically accepted by our readers but I did not attempt (perhaps did not dare) to publish it in English. After Professor Wheeler authorized the English version in January 1977, the Czech translation was published in \v{C}eskoslovensk\'{y} \v{c}asopis pro fyziku A 28 364-374 (1978) and soon afterwards the Polish translation appeared in Post\c{e}py fizyky 29 523-534 (1978). The last time I met John Wheeler was at the 12th Pacifc Coast Gravity Meeting and Karel Kucha\v{r} fest in March 1996 in Salt Lake City. He recalled the interview and, in fact, expressed a wish to continue with some questions which later occupied his mind primarily---the problem of the quantum and its relation to reality. He suggested the interview could be published in \emph{Physics Today}. But I never asked more questions and I never tried to send the interview to an English journal. After John Wheeler's recent death it occurred to me that it would now be appropriate to publish the original interview from 1976 so that it would not be lost to English readers; and so, despite being more than 30 years old, the interview appears here. John Wheeler would now surely add more about black holes in nuclei of galaxies, not mentioning just Cygnus X-1, when discussing cosmology he would undoubtedly address the problem of dark energy etc. However, in the conversation about Einstein and Bohr, about the need for choosing appropriate names, or about the relation of science and philosophy and art, he would probably give answers as he did more than 30 years ago.

\vspace{1em}

\emph{In the last twenty years, Professor Wheeler, you have changed significantly the field of your primary interest: after having done much influential work in nuclear physics and related regions you turned to general relativity and cosmology, created a school, developed many new ideas and attitudes---became a world leader in general relativity. Would you, please, recall the impulses and motivations which led you to get into a new field?}

Historically, I had been concerned with trying to find a simple description of the force between particles both in nuclear physics and more generally. That had led to the work of Feynman and myself on action at the distance as a way of describing the electromagnetic coupling between one particle and another. And from that I had gone on to start to work on finding the equivalent description in gravitation theory, in gravity. In electromagnetism, action at the distance means two particles coupled directly rather than making any mention of the field of force between them. In the case of gravitation, this would mean that one would talk of a coupling between one particle and another with no mention of the space and time between them. If electromagnetic interaction at a distance sweeps out the field from between the particles, then gravitational action at a distance should sweep out space and time from between the particles. 

I had started on that but then my life had been very much affected by getting involved in national defense questions. And then, when I came back again after three years to gravitation, I started giving a course in relativity and that brought to my attention how many great questions there are in relativity and, particularly, the question about gravitational collapse. I had got interested in the question of trying to find a simple model of gravitational collapse, a case where one would be free of all the issues of matter and its equation of state and its density, where one could talk in a very simple language. And it seemed it would be the most simple description, most simple example, most simple problem of gravitational collapse to talk of a star which was made entirely of photons---then all the gravitational attraction will arise from something one understands well. So this led to the idea of a ``geon'', an object composed entirely of radiation but appearing from the outside as a concentration of mass. Well, I got in on many problems since that time in 1953 when I first started to teach general relativity, but always at the centre of my attention has been this question of gravitational collapse because it first made one recognize that one was moving into a new field.
 
You could say why should one get into a new field like that. I can remember coming back from the first hydrogen bomb test in the Pacific at Eniwetok on an airplane. And then I came to Honolulu, was there a few hours, and at that moment a tidal wave came---which had nothing to do with the hydrogen bomb, it had to do with the great earthquake in Kamchatka. But I thought how small the efforts of man are. Even the biggest explosion that the United States ever made, which my group at Princeton did the design work for---working for and with Los Alamos---even that greatest explosion was a thousand times smaller in energy than the energy of a hurricane, a thousand times smaller than the energy of an earthquake. And then I could not help thinking of the big bang and the expansion of the universe so much larger, and as one flies over the infinite distances of the Pacific ocean suspended between the heaven above and the ocean below, one feels he is somewhere in the space between the stars like man's position in the universe, and one realizes what great mysteries surround us. One asks himself, how can I get at the heart of this mystery of the universe. And I've never seen a more central place to get at the heart of this mystery than gravitational collapse.

\vspace{1em}

\emph{Yes, I remember how Kip Thorne, in his article in ``Magic without Magic'', recalls the time in 1962 when he, as a new graduate student in Princeton, for the first time entered your office and you began immediately to discuss the many unsolved aspects of gravitational collapse. And as he writes, ``John Wheeler alone was grinding the axe of gravitational collapse (to put in crude terms what he did so elegantly) before quasars, before pulsars, before singularity theorems\ldots'' By now we know most vividly how the question of gravitational collapse has been followed in recent years. What today is your opinion on future developments, what kind of methods do you now expect to be of importance in solving the problem of gravitational collapse? In particular, do you expect that it will, like nuclear physics appears today, be predominated by sophisticated numerical methods in near future?}

On this question of gravitational collapse we have today, of course, an enormous amount of information, and you have been one of the leaders in discussing the question of the field around a collapsed body. We now have also beautiful theorems about how standard the conditions are on the outside of the collapsed body. But all the irregularities, all the disturbances, all the hydrodynamics, all the magnetic fields, all the turbulence, all the entropy that goes into the horizon and gets hidden behind the horizon in gravitational collapse, leaving these beautifully standard conditions outside surely must show up inside of the horizon, so it seems to me. That inside must be extremely interesting, extremely full of violence and turbulence. And I can well believe that we can find methods and will find methods and must find methods to see what goes on there---whether they will be numerical or whether they will be in terms of what one calls a qualitative theory of differential equations, or some mixture of the two, only the future will tell. I suppose that the main reason for interest in this is to see what kind of physics goes on there. If we say a black hole has the wonderful feature of giving us an example of gravitational collapse without our having either to go back to the big bang at the start of the universe or to go forward to the big stop at the end of the universe, then everybody will feel certain vividness about dealing with the black hole.

\vspace{1em}

\emph{You mentioned the ``big stop'' at the end of the evolution of the universe. It is well known that you are a great advocate of the idea of a closed universe in which a recollapse into the singularity occurs, in contrast with an open universe which keeps expanding forever. What are your reasons for the belief in the closed model of the universe?}

This wonderful issue of the open universe compared to the closed universe is most lively at the present time. I can believe that debate about it and analysis of it will grow in intensity in the coming year, in the coming several years. Einstein long ago, of course, was led into general relativity not least by his idea---going back to Ernst Mach---that the inertia of one particle here and now arises from its interaction with other particles elsewhere in the universe. Later on, in his famous book, The Meaning of Relativity, p.\,150, he talks about his reasons for still believing in a closed universe. Closure would mean a finite number of particles in the universe for given particle to interact with.

Today, of course, there is another reason that one has to think of a closed universe instead of an open universe: there is no natural way to define the boundary conditions for an open universe. One might at first think the most natural boundary condition for the universe is asymptotic flatness. But in a universe that goes asymptotically flat one has no way to define what flat is in the framework of a modern quantum theory. The metric is really oscillating and fluctuating everywhere. No matter how great the distance to which one goes one never comes to a distance so great that space becomes flat. Therefore ``asymptotic flatness'' is a physically impossible boundary condition. No alternative boundary condition has ever been proposed for an open universe that does not run into the same difficulty. Closure is the only boundary condition we know that is at the same time mathematically well-defined and physically reasonable.

But still we have to recognize that the universe is not something that we necessarily can be confident in making theories about. We have to be open to the possibility that the evidence might some day compel us to think of the universe as open. In that case I suppose that we will be forced to say that as time goes on in the history of the universe we are all the time getting information from greater and greater distances in the universe and new parts of the universe will forever be swimming up into our horizon and sending signals to us. 

But I do believe it is worthwhile remembering back to 1953. That was the time when it looked as if one were in great difficulty with Einstein's idea of a universe with its expansion slowing down with time. The astrophysical evidence at the time pointed to an expansion of the universe that speeded up with time. Imaginative investigators put forward all kinds of theories like ``the theory of continuous creation'' and ``the steady state universe''. Every one of those theories meant giving up Einstein's simple ideas. In the end it turned out that they were all wrong---that Einstein's original idea was correct that the expansion of the universe is slowing down. The trouble was only that the astrophysical data on distances to the other galaxies had been wrong by a factor of six. 

That is the story of one difficulty with the idea of a closed universe---a difficulty that turned out not to be a difficulty. Well, today what is the difficulty with the idea of the closed universe? It's primarily that we do not see enough matter around; we appear to be short by a factor of something like 30 from having enough matter to curve up the universe into closure. However, today our colleagues in the world of astrophysics are beginning to tell us that there is much more matter in the universe than we had realized a few years ago. They find evidence that typical galaxies weigh somewhere between 3 and 20 times as much as one had first believed.

\vspace{1em}

\emph{It is, of course, a wonderful outcome of recent years that astrophysics and, in particular, general relativity and cosmology have left the ivory tower of theoretical speculations and have become directly connected with experiment. What, according to your opinion, will be the role of future experiments, using more and more advanced technology?}

Yes, I believe that everybody would agree that astrophysics has developed absolutely spectacularly in the last five and ten years. We have new telescopes, we have X-ray astronomy, we are getting infrared astronomy, radio astronomy continues its marvelous output, and we look forward to gravitational wave astronomy and neutrino astronomy. Fifteen years ago who would have believed that we could hope to know anything nearly as much as we do today about the far away and long ago! What an enormous development!

On relativity, too, observation contributes more than ever, and especially in connection with gravitational collapse. First came neutron stars, and then our present incomplete but partly convincing evidence that the X-ray source Cygnus X-1 is a black hole. It focused attention more than ever on gravitational collapse and on the search for gravitational radiation which is going on so actively now. If gravitational wave detectors bear out their promise---not fifteen detectors of low sensitivity but three detectors of high sensitivity---if they turn out to work well and give occasional evidence of events, they will achieve two goals at once: (1) give us additional evidence that relativity is right and (2) yield new and direct information about what goes on in the interior of distant stars.

\vspace{1em}

\emph{Allow me now to go over from general relativity to its creator himself. In fact, you are one of very few physicists who were not only closely related to Albert Einstein, but you were also one of the nearest collaborators of Niels Bohr. Could you, please, characterize by what have you been most influenced when having been in contact with these two greatest architects of modern physics?}

It has been a wonderful inspiration to know both men. I first came to know Einstein in 1934 on my first visit to Princeton, very shortly after he had come to the United States. And then in 1953, I remember when I first started to teach relativity that although it was only 18 months before his death, he was kind enough to invite me to bring my students around to his house for tea. So we sat around the dining-room table and his secretary Helen Dukas and his stepdaughter Margot Einstein brought tea and students asked Einstein questions. One of them: ``Professor Einstein, what do you think about the nature of electricity?''---and he told about his thoughts over the years about electricity. And another one: ``Professor Einstein, do you agree with the idea of an expanding universe?''---and, of course, he did. And another student: ``Professor Einstein, you had so much to do with the quantum theory and why don't you agree with the quantum theory?'' And then Einstein said again that he did so often in his famous words: ``I do not believe that God plays dice.'' And finally one student got up his courage and he said: ``Professor Einstein, when you are no longer living, what will happen to this house?'' And Einstein gave a great big laugh, he threw up his hands and with his hearty voice he spoke with a childlike simplicity and smile on his face and bright eyes---and his choice of words was always so careful and so beautiful: ``This house will never become a place of pilgrimage where the pilgrims come to look at the bones of the saint.'' And so it is today. The tourist buses drive up in the front of his house and people get out and photograph the outside but they do not go in the inside.
 
And then Bohr, Bohr the greatest leader of physics and father-figure of all physicists. I went to Copenhagen and I can remember, as a student applying for a fellowship, the words I put down in my application for the fellowship---why I wanted to go. That was in 1934, very early 1934. Why did I want to go to work with Bohr in Copenhagen?---It was because ``he has the power to see further ahead in physics than any other man alive''. From my arrival in September I saw his great gift to think deeply in nuclear physics. There in Copenhagen in the spring of 1935 Christian M\o ller, fresh back from Rome, reported Fermi's results on the capture of slow neutrons. Bohr immediately became terribly concerned, interrupted, walked up and down, talked and talked, and as he talked one could see the liquid drop model of the nucleus taking shape right there before one's eyes. For him no physics was of any interest unless it yielded some paradox or some beautiful way of seeing things simply. I do not remember anyone at Bohr's institute who ever succeeded in finishing a seminar talk, even though he was the invited speaker. He might be able to speak five minutes, he might be able to speak fifteen minutes, but soon Bohr would take over and would use the whole time discussing the meaning of the speaker's results and what they proved and what did not prove.

I became involved with Bohr on nuclear fission at the time when he brought word of the discovery of fission to the United States on the sixteenth of January in 1939. I was down at the waterfront pier in New York and I had hardly said ``Hello'' when he took me aside and started to tell me that on this very ship just before he left Copenhagen he had been told about the discovery of Hahn and Strassmann. So we dropped everything else and started to work on fission. I can remember rushing---we worked at night as well as in the day---rushing up the steps in the library---from my office to the library--- to look at the dictionary to see whether there is a better word than ``fission''. ``Fission'' had an unfortunate property. The noun is all right but there is no good verb. A nucleus ``fissions'' is not a very nice verb, but we stayed with ``fission'' in spite of that.

During the war, I met Bohr in Washington at the time he was dividing his time between Los Alamos and Washington after he had escaped from Denmark in a small boat over the sea into Sweden. He told me confidentially about his discussions with President Roosevelt about the future of nuclear energy. He told me about his efforts to work out some control of nuclear energy after the war. He said, ``It may seem strange, how can such a man as I speak to the president of the greatest country of the World at the time of the greatest war in the history of the world. But'', he said, ``I put it to him as man to man simply in terms of what the problem is and what other possibility is there than this.'' Bohr made a great impression on Roosevelt and they had several discussions. The last speech that Roosevelt wrote---he died while he was still working on that speech---had in it some words, quoted by Roosevelt from Thomas Jefferson, about how scientists serve as the most important means of communication and bringing peace between the different countries of the world. It was enormously impressive to me to see Bohr's courage in facing up to what the great questions are. I can remember his particularly saying to me at one time: ``I must seem always to you like an amateur. But I am always an amateur.'' Of course, it's a very modest way to say that one is a pioneer, an explorer. If you are working on something new, then you are necessarily an amateur.

To me the debate between Bohr and Einstein over the years is the greatest debate in all the history of human thought. I can't think of any greater men debating any deeper issue. It took place for a number of years in Europe and then for a number of years in America. Unfortunately, the artists of the West seem not to be so much aware of science. But in 1971, on an earlier visit to Moscow, I visited a studio in the basement of an apartment house where two sculptors were making sculptures of artists and poets, and great thinkers, and great scientists. There was a sculpture of Bohr and Einstein debating. It was wonderful to see that. But I did not tell the sculptors about one of the times when Bohr came to visit the house of Einstein that I just mentioned. He went up the stairs to the second floor where Einstein's study was and---it was a terribly hot day---he found Einstein lying on the sofa with no one piece of clothing on. Well, they continued the debate in that frame of reference. The sculptors did not know that.

The debate concerned what to my mind concerns the deepest, the most mysterious, the most challenging idea in all of physics, the quantum principle, the overarching principle of twentieth century physics. As you know, while Einstein was still in Europe the debate focused on Einstein's belief that quantum theory was inconsistent. He did not only talk. He tried to give a proof that the uncertainty principle is logically inconsistent. At the famous Solvay Congress of October 1930 Einstein confronted Bohr with his idealized experiment. How dramatic it was when Bohr turned the tables and used Einstein's own general relativity to prove that Einstein's scheme would not work! After Einstein came to the United States, he gave up trying to prove that quantum theory is inconsistent. He now tried to prove that the quantum theory is incompatible with any reasonable idea of reality. His efforts led to the famous Einstein--Rosen--Podolsky ``paradox'' which at the hands of Bohr and Bell and others has brought us so much understanding.

We shall have to let science unroll through the years ahead before we can look back and know which is the greater man because we know each new generation has new insight on history. But to me the two men had so much in common. They were so happy talking with each other. They were always concerned with the deepest problems, not only problems of physics but the deepest problems of mankind.

Einstein preferred to work in isolation. Bohr was greatly stimulated by having collaborators to talk and argue with.

Bohr was deeply convinced that cooperation in scientific research offers more than any other policy the opportunity for bringing about close contacts and common understanding between nations. He believed that the development of science plays the most important role in tieing together different cultures. We remember his open letter to the United Nations 1950. His idea of an open society failed to have much influence at that time, but today we feel it is making more headway: not to say that this system is better or that system is better but let everyone visit where one will and draw one's own conclusions.

\vspace{1em}

\emph{As you indicated, Niels Bohr created one of the most influential schools of modern physics in Copenhagen. But it is well known that you have also educated many leading physicists both in nuclear physics and in general relativity in Princeton. What are your basic ``rules of interaction'' with students?}

Isn't there a mistake in your question? I am sure that it is really the students who educate me! We all know that the real reason universities have students is in order to educate the professors. But in order to be educated by one's students one has to put good questions to them. One always tries out questions on one's students. There will be some questions no student gets interested in, and if after a while no student gets interested in a certain question, then you know that question is not very good and you throw it away. But if there are questions that the students get interested in, then they start to tell you new things and keep you asking new questions, and pretty soon you have learned a good deal. One feels very happy when a student gets to feel, as one does himself, that the whole world of science is like a gigantic pie, and you can cut in the pie anywhere you wish, and take out a slice and eat it.

But the wonderful thing about it is that one finds that with the right kind of students, they are not interested in small things, they want to do things that are important. But, of course, it's important also not to make everything too important because one has to keep in touch with reality. It's very nice to have some reality. I brought here with me a piece of reality. This is a piece of concrete---but it's very interesting because the reinforcing in the concrete is made with steel rods not as big as ordinary steel in reinforced steel, but in the form of hairs. Each piece is about 3 cm long, and is big around as a pin. One mixes in these ``hairs'' or ``pins'' with the sand and gravel and cement and water at the time one makes the concrete. I am very interested in this type of concrete because I think it is going to be an entirely new building material. I try to give some encouragement to the people who work on this.

\vspace{1em}

\emph{Yet, a short question concerning students. Do you prefer to speak with them individually or, rather, do you organize a number of regular informal seminars?}

I personally learn more by talking to the individual student.

\vspace{1em}

\emph{You invented not only many new ideas but also new names to new ideas. ``Black hole'' is an example of such a name which is now accepted all over the world. But there are other examples: ``moderator'', ``buckling'', ``big stop'', ``charge without charge'', ``mass without mass'', etc. Why should not there be a new idea without a new name?}

Mark Twain used to say, ``The difference between the right word and the nearly right word is the difference between lightning and a lightning-bug.'' It is an old idea among mankind that if you can name something you can somehow get control of it. And even the doctors have convinced us that we should pay them for giving names to our diseases.

But giving the right name is a part of a larger way of grasping ideas, it seems to me, which I noticed so much not only in Bohr and Einstein but also in Pauli. Pauli once put it in these words: ``What is the wit of it?'' He meant, what is the central point of it. If one could not state the central point in two or three words or in one sentence, then one did not really understand it. What a stimulus to thought it is to be forced to look into an idea so deeply that one can state it in a simple phrase!

In my country there is a great deal of interest in advertising. It is sometimes looked down upon a little but people nevertheless take an amused interest in it. There is one very famous advertising man called Ogilvy who wrote the book called ``The confessions of an advertising man''. He tells there about how one of his most important duties in helping to advertise a company or its product is for him to find out---or to make that company to think out for itself---what it stands for. The most famous example is the Avis company which rents automobiles. He forced them to think out for themselves what did they have to offer when there was a bigger company---the Hertz company---that also offered automobiles for rent. And finally he and the Avis people found it: ``We try harder''. ``We try harder'' became the advertising slogan of the Avis company. It had and still has a great psychological effect on the people who work for the company. They indeed felt: ``We try harder''. So, I think, the right phrase has a magic about it.

When I watched in our meeting today people making notes of what the speakers say, I finally concluded that every one of us wants to be a magician and he thinks that somehow by getting these magic formulas he can do what Merlin, the magician, used to do in the old days with his magic incantations and his magic formulas. But part of the magic is finding a simple word.

\vspace{1em}

\emph{Thank you for bringing us nearer to understanding of ``Magic without magic''. Of course, conservatives probably do not welcome such new names as ``black hole'' or ``mass without mass''. But let me ask you about a better type of conservatism. Yesterday I have seen a remark in a little book called ``Written into the clouds'' by Josef \v{C}apek (the painter, writer and the man as deep as his more known brother Karel) that ``the only apology for a cultural conservatism is concern that nothing that we have learned about the world should be diminished and nothing should be lost''. Similarly one may perhaps look on a conservatism in science. What do you think about scientific conservatism, what do you think about fashion in science?}

For the two-hundredth anniversary of the independence of the United States, the National Academy of Sciences a few weeks ago had a special meeting on the future of science and I had to give the first talk there on the future of physics. In trying to find a single phrase, a single idea that would summarize more clearly than anything else the future of physics as I see it, I found nothing better than to compare physics with life. Life evolves in many different directions to fill as our friends in the world of biology call it every ecological niche. Some people work on solid state, some people work in a conservative way and some people are focused on applying physics to medicine. My feeling is that there is as much room for different kinds of physics as there is room for different kinds of people. Everybody has a contribution to make.

\vspace{1em}

\emph{Conservatism in art led me to ask you about conservatism in science. But I would like to ask you also quite generally: what do you feel about the relationship between science and art? I remember that it was Richard Feynman who mentioned elsewhere how much deeper is the view at a seashore if one adds to the view of an artist a knowledge of hydrodynamics and molecular physics. You draw such beautiful pictures to explain your ideas in physics that also because of this I yearn to ask you such a question.}

It is a deep question you ask and an interesting question and I remember so well the words of one artist who was kind enough to give me art lessons in Paris in 1949. I went twice a week to him for drawing. He told me how he had got his education at the \'{E}cole des Beaux Arts in Paris. He said that his fellow students there were so well trained in observing things carefully and accurately, to get the truth, that they understood him better than his own father and mother understood him. This made a great impression on me---this concern for accuracy and truth.

But to me also it was very interesting the idea that in art you are trying to distill out of the situation some central thing and find out what that central thing really is and capture it in its naked essence, free of all complications. And that to me is what is so impressive in science. There, too, we are trying to do this all the time: capture the naked essence of the situation in the very simplest terms. So, to me there is a very great similarity between the two: the search for truth and the search for the absolutely central point.

But certainly there is also a difference. A work of art only really comes alive if it produces some resonance in the hearts of the people that look at it. Something may be a wonderful work of art but if the people are wrong people to look at it, it has no effect. It is tied, therefore, to the human heart in a way much closer than science is. It is true that science is a human activity, and it is a collaborative activity, and it is true that if someone does a piece of work and nobody pays attention to it, then it has no effect. But in the case of science you could say that there is a kind of democracy about it. The steps in a proof are democratically open for everybody, or for every qualified person, to check for himself. Or an experiment is democratically open for anybody to check for himself if only he knows how to do experiments. In the case of art, well, I suppose, one would say there, too, that it is democratically open to anybody to resonate to it but it does not have the same compulsion about it. In the case of the proof---there is the proof, in the case of the experiment---there is the experiment. You will come out with ``yes'' or ``no'' at the end of it. But in the case of the work of art it is not ``yes'' or ``no'', it is resonance. 

\vspace{1em}

\emph{You are not only a member of the American Academy of Arts and Sciences, which combines both creative activities you spoke now so beautifully about, but also a member of the American Philosophical Society. So, I would like to ask you yet, what is your opinion on the relation between science and philosophy. Even with such an outstanding place concerned with fundamental aspects of physics as Caltech, for example, one feels that at present there is no immediate fruitful interaction between science and the philosophy of science. But, perhaps, Caltech is a more ``pragmatic'' school than Princeton.}

Clemenceau, the Prime Minister of France in World War One, said that war is too important to be left to the generals. He took control of the situation. And one could say that philosophy is too important to science to be left to the philosophers. But there are two extreme views. There is the view of one man who describes the philosophy of science as a tin can which is tied by string behind the automobile of science. And as science goes quietly ahead, this tin can rattles on the street and it is what makes all the noise. That is one view. But the other view is much deeper. Thomas Mann, in his lecture celebrating the eightieth birthday of Sigmund Freud, said: ``Science never makes an advance until philosophy authorizes and encourages it to do so''\ldots

Well, you can have your choice between those two views! 
\vspace{1em}

\emph{It was a wonderful experience to listen to your words. Thank you very much. Ji\v{r}\'{i} Bi\v{c}\'{a}k.}

\end{document}